\documentclass[11pt,twoside,letterpaper]{article} 
\usepackage{times,fancyhdr,amssymb}
\usepackage[dvips]{graphicx}
\makeatletter 
\def\cleardoublepage{\clearpage\if@twoside \ifodd\c@page\else%
    \hbox{}%
    \thispagestyle{empty}%
    \newpage%
    \if@twocolumn\hbox{}\newpage\fi\fi\fi} 
\makeatother 

\newcommand{\etal}{{\it et al.}}
\def\figurename{Figure}
\makeatletter
\renewcommand{\fnum@figure}[1]{\figurename~\thefigure.}
\makeatother
\def\tablename{Table}
\makeatletter
\renewcommand{\fnum@table}[1]{\tablename~\thetable.}
\makeatother
\graphicspath{{figs/}}
\begin{document}
\title{
{\begin{flushleft}
\vskip 0.45in
{\normalsize\bfseries\textit{Chapter~1}}
\end{flushleft}
\vskip 0.45in
%
%
%
%
\bfseries\scshape Nuclear pasta in supernovae and neutron stars}}
\author{\bfseries\itshape Gentaro Watanabe$^{1,\, 2,\, 3\,}$\thanks{E-mail gentaro\_watanabe@apctp.org}\quad and 
Toshiki Maruyama$^4$\thanks{E-mail maruyama.toshiki@jaea.go.jp}\\
$^1$Asia Pacific Center for Theoretical Physics (APCTP)\\ POSTECH,
San 31, Hyoja-dong, Nam-gu, Pohang, Gyeongbuk 790-784, Korea\\
$^2$Department of Physics, POSTECH\\
San 31, Hyoja-dong, Nam-gu, Pohang, Gyeongbuk 790-784, Korea\\
$^3$Nishina Center, RIKEN, 2-1 Hirosawa, Wako, 
Saitama 351-0198, Japan\\
$^4$ASRC, Japan Atomic Energy Agency, Tokai, Ibaraki 319-1195, Japan
}
\date{}
\maketitle
\thispagestyle{empty}
\setcounter{page}{1}
\begin{abstract}
In supernova cores and neutron star crusts, nuclei with exotic shapes
such as rod-like and slab-like nuclei are expected to exist. These
nuclei are collectively called nuclear ``pasta''.
For the past decades, existence
of the pasta phases in the equilibrium state has been studied using
various methods. Recently, the formation process of the pasta phases,
which has been a long-standing problem, has
been unveiled using molecular dynamics simulations. In this review, we
first provide the astrophysical background of supernovae and neutron
stars and overview the history of the study of the pasta phases. We then
focus on the recent study on the formation process of the pasta phases. 
Finally, we discuss future important issues related to the pasta phases: 
their astrophysical evidence and consequences.
\end{abstract}
\thispagestyle{fancy}
\fancyhead{}
\fancyhead[L]{In: Neutron Star Crust \\ 
Editors: C.A. Bertulani and J. Piekarewicz, pp. {\thepage-\pageref{lastpage-01}}} 
\fancyhead[R]{ISBN 0000000000  \\
\copyright~2012 Nova Science Publishers, Inc.}
\fancyfoot{}
\renewcommand{\headrulewidth}{0pt}
\vspace{2in}
\vspace{.08in} \noindent \textbf{}
%
\pagestyle{fancy}
\fancyhead{}
\fancyhead[EC]{G. Watanabe and T. Maruyama}
\fancyhead[EL,OR]{\thepage}
\fancyhead[OC]{Nuclear pasta in supernovae and neutron stars}
\fancyfoot{}
\renewcommand\headrulewidth{0.5pt} 
%
\newcommand{\lsim}{\lesssim}
\section{Introduction\label{sect_intro}}

\begin{figure}[tbp]
\begin{center}
\resizebox{11cm}{!}
{\includegraphics{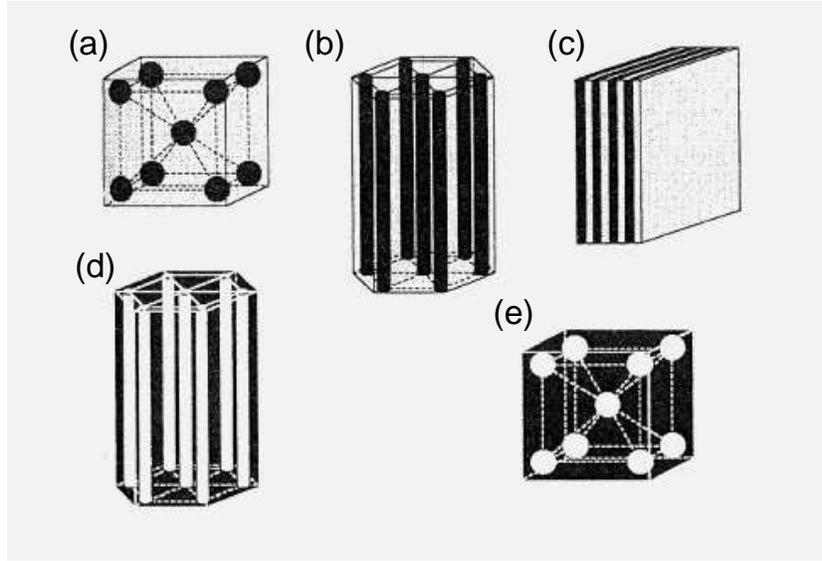}}
\caption{\label{fig_pasta} \quad
  Schematic pictures of nuclear ``pasta''.  The black regions show
  the liquid phase of nuclear matter, where protons and neutrons coexist.
  The gray regions show the gas phase, which is almost free of protons.
  Nuclear shape changes in the sequence from (a) to (e) with increasing
  density.  Figure courtesy of K. Oyamatsu \cite{Oyamatsu93}.
  }
\end{center}
\end{figure}

In terrestrial matter, atomic nuclei are roughly spherical.  
However, this common wisdom does not necessarily hold
for matter in supernovae and neutron stars.
There, density of matter is very high and can reach the value
inside the atomic nuclei themselves, i.e., the normal nuclear density
$\rho_0 \simeq 0.165$ nucleons fm$^{-3}$
corresponding to $\simeq 3\times 10^{14}$ g cm$^{-3}$.
In such a high-density environment, nuclei will adopt 
various shapes such as spaghetti-like rods and lasagna-like slabs
\cite{Ravenhall,Hashimoto} (see Fig.\ \ref{fig_pasta}).
These non-spherical nuclei are collectively called nuclear ``pasta''
and the phases with these nuclei as ``pasta'' phases
\footnote{According to Bethe \cite{Bethe90}, the term ``nuclear pasta''
was coined by Cooperstein and Baron \cite{Cooperstein90}.  However,
Ravenhall may be the first person who has coined this term \cite{Pethick11}.}.

In the liquid drop picture of atomic nuclei, energy of nucleus
consists of bulk energy of nuclear matter region, surface energy 
of the nucleus, and the Coulomb energy between protons in the nucleus.
Among these three contributions, the latter two depend on the 
nuclear shape: the surface energy favors a spherical nucleus
and the electrical repulsion energy between protons tends to make
the nucleus deform.
At low densities far below the normal nuclear density $\rho_0$,
the effect of the surface energy is predominant, and thus the shape of
nuclei is spherical.
However, when the density of matter reaches around $\lesssim\rho_0/2$ and 
nuclei are closely packed, 
the effect of the electrostatic energy becomes comparable to 
that of the surface energy and
the total energy of the system can be reduced by
combining nearest neighbor nuclei and decreasing 
the total surface area of nuclei.
Consequently, the energetically favorable configuration is expected to have
remarkable structures of the pasta phases:
the nuclear matter region (i.e., the liquid phase region)
is divided into periodically arranged parts of roughly
spherical (a), rod-like (b) or slab-like (c) shape, 
embedded in the gas phase and in a roughly uniform electron gas
\footnote{In the relevant density region, the screening effect of electrons
is negligible \cite{Pethick95,screening,Maruyama05}.}.
Besides, there can be phases in which nuclei are turned inside out,
with cylindrical (d) or spherical (e) bubbles of the gas phase
in the liquid phase.

In the present article, we provide an overview of the recent
progress of the study on the nuclear pasta phases.
The structure of this article is as follows.
In the remaining part of the present section, 
we provide a brief explanation about astrophysical background 
relevant to the pasta phases.
In Section 2, we overview the study of the
pasta phases so far, and then in Section 3
we explain the molecular dynamics approach of nucleon many-body systems.
In Section 4,
we explain the recent study 
about the dynamical formation of the pasta phases.
Finally, we discuss the future prospects of the study of the pasta phases,
especially focusing on their astrophysical consequences.

\subsection{Astrophysical background \label{sect_astro}}

\subsubsection{Core collapse supernovae \label{sect_sn}}

Core collapse supernova explosion is the final stage of the evolution
of massive stars whose main-sequence mass is $\sim 8-30 M_{\odot}$
(for reviews, see, e.g., Refs.\ \cite{Bethe90,Suzuki94,Kotake06,Janka07}).
In such stars, as the nuclear burning proceeds, 
iron core is formed, grows in mass, and finally becomes
unstable to gravitational collapse triggered by 
electron capture on iron nuclei, 
$^{56}\mathrm{Fe}+e^- \rightarrow\ ^{56}\mathrm{Mn}+\nu_e$.
This reaction leads to a reduction of the degeneracy pressure of electrons,
which is a major pressure source supporting the iron core.
Besides, when the temperature exceeds about $5\times 10^9$ K,
thermal energy generated by the core contraction is 
consumed by an endothermic reaction of the photodissociation of iron nuclei,
$\gamma +\ ^{56}\mathrm{Fe} \rightarrow 13\ ^{4}\mathrm{He} + 4n
 -124.4 \mathrm{MeV}$.  This process acts as a positive feedback
and accelerates the collapse.

As the gravitational collapse proceeds, density in the core becomes so high
that even weakly interacting neutrinos are temporarily trapped in the core
for $\sim 10$ msec \cite{Sato75}.
Namely, when the average density of the core exceeds $\sim 10^{11}$ g cm$^{-3}$,
diffusion time scale $\tau_{\rm diff}$ of neutrinos becomes comparable to
or larger than dynamical time scale $\tau_{\rm dyn}$ of the collapse.
Here, $\tau_{\rm diff}$ is given by the random-walk relation: 
$\tau_{\rm diff}\sim R^2/cl_\nu$,
and $\tau_{\rm dyn}$ by the free-fall time scale: 
$\tau_{\rm dyn}\sim 1/\sqrt{G\rho_{\rm core}}
\sim 4\times 10^{-3} (\rho/10^{12} \mathrm{g\, cm}^{-3})^{-1/2}\, \mathrm{sec}$,
where $l_\nu$ is the mean free path of neutrinos,
$M\sim 1 M_\odot$ and $R$ are the mass and the radius of the core, 
$\rho_{\rm core}\sim M/(4\pi R^3/3)$ is the average density of the core,
$c$ is the speed of light,
and $G$ is the gravitational constant.

\begin{figure}[tbp]
\begin{center}
\rotatebox{0}{
\resizebox{8.5cm}{!}
{\includegraphics{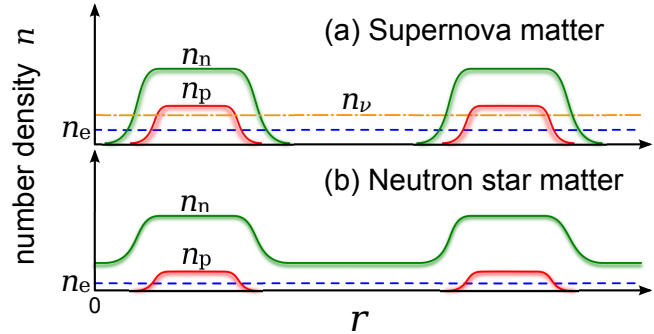}}}
\caption{\label{fig_wscell}
 Schematic plot of the density profiles in
 (a) supernova matter and (b) neutron star matter 
 at subnuclear densities. 
 Here $n_n$, $n_p$, $n_e$ and $n_{\nu}$ are
 the number densities of neutrons, protons, electrons, and neutrinos
 respectively.
  }
\end{center}
\end{figure}

While neutrinos are trapped in the core, 
lepton (i.e., electron plus neutrino) number per nucleon (lepton fraction)
is fixed around 0.3 -- 0.4.  
Trapped neutrinos become degenerate and form a Fermi sea, which
suppresses the electron capture reaction.
Therefore, matter in supernova cores [i.e., supernova matter]
is not so neutron rich and the typical value of the proton fraction $x$
(ratio of the number of protons to the number of nucleons) is $\simeq 0.3$. 
In Fig.\ \ref{fig_wscell}(a), we show the schematic picture of the
particle distributions in supernova matter:
almost all protons and neutrons are bound in nuclei and
charge-neutralizing electrons and trapped neutrinos are
distributed uniformly in space.

After $O(100)$ msec from the onset of the collapse,
the central density of the core reaches
the normal nuclear density $\rho_0$.
Then, the short-range repulsion in nucleon-nucleon interaction 
starts to act and the equation of state (EOS) suddenly becomes stiff.
Due to this hardening, pressure becomes sufficiently high to halt
the collapse, causing the inner region of the core to bounce.
The outer region of the core continues to fall towards the center at
supersonic velocities. Consequently, the bouncing inner core drives
a shock wave outward into the infalling outer core.
After the core bounce, temperature in the inner core becomes so high
($\gtrsim 10 - 20$MeV) that nuclei melt.

In the process of the collapse, nuclear pasta phases
are expected to be formed 
in the region corresponding to subnuclear densities.
They would exist for $O(10 - 100)$ msec until they melt
after the bounce of the core.
The total amount of the pasta phases could be more than 20\% 
of the total mass of the supernova core just before the bounce \cite{opacity}.

The shock wave propagates outward in the outer core.
However, the shock wave is weakened by the following processes:
1) Energy dissipation
due to dissociation of nuclei into free nucleons by the shock itself and
2) the pressure reduction behind the shock front due to
the emission of neutrinos created by the electron capture
on protons coming from the dissociated nuclei.
Consequently, the shock wave stalls in the outer core.
On the other hand, thermal neutrinos are emitted from the hot inner core
(proto-neutron star)
and they heat the material in the outer core
(referred to as neutrino heating). 
If the energy transfer from neutrinos to the matter behind the shock front 
is large enough, the stalled shock can be revived and
reach the surface of the outer core by $\sim 1$ sec after the bounce. 
The shock wave can propagate outside the core without obstacles
and finally blow off the outer layer of the star 
including a part of the outer core after between 
a few hours and a few days; 
this explosive phenomenon is a core collapse supernova explosion.

\subsubsection{Neutron stars \label{sect_ns}}

\begin{figure}[t]
\begin{center}
\rotatebox{0}{
\resizebox{11cm}{!}
{\includegraphics{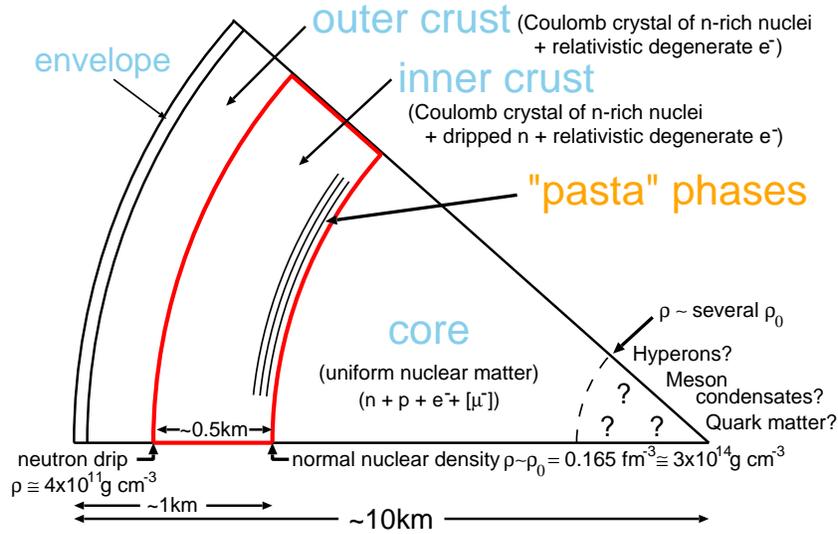}}}
\caption{\label{fig_ns}
  Schematic picture of the cross section of a neutron star.
  This figure is taken from Ref.\ \cite{Watanabe_review}.
  }
\end{center}
\end{figure}

While the shock wave propagates in the outer core, 
the proto-neutron star contracts from $\sim 100$ km to $\sim 10$ km.
Then, in the time scale of $\sim 10$ sec, thermal neutrinos trapped 
in the proto-neutron star diffuse out.  
As results, temperature of the proto-neutron star decreases
and matter in the proto-neutron star becomes neutron rich: 
the proto-neutron star becomes a neutron star.
When the temperature decreases to $\lesssim 10$ MeV,
nucleons in the gas phase 
start to cluster to form nuclei below $\sim \rho_0$.

Neutron stars are dense and compact objects supported by the degeneracy
pressure of neutrons 
(see, e.g., Refs.\ \cite{Pethick95,Chamel08} for reviews). 
The typical mass and the radius of a neutron star are
$\simeq 1.4 M_{\odot}$ and $\simeq 10 \mathrm{km}$, respectively 
\footnote{Recently, a heavy neutron star with mass $1.97 M_\odot$ 
has been discovered \cite{ns2solar}.}.
In Fig.\ \ref{fig_ns}, we show a schematic picture of the structure
of a neutron star.
Neutron stars roughly consist of two main regions:
crust, which is made of crystalline lattices of neutron-rich nuclei,
and core, which is made of liquid of nuclear matter
(Deep inside the core corresponding to several times normal nuclear density 
is very uncertain and various hadronic phases and quark matter might exist.
These are beyond the scope of the present article.).

In the crust of neutron stars, nuclei in 
deeper region (i.e., at higher densities) are more neutron rich
due to larger chemical potential of electrons, which promotes
electron capture.
At a density 
$\rho_{\rm drip}\simeq 4 \times 10^{11} \mathrm{g}\,\mathrm{cm^{-3}}$, 
nuclei become so neutron rich that all neutrons cannot be bound 
in nuclei and they start to drip out.
This neutron drip density divides the crust into two regions:
outer crust $(\rho<\rho_{\rm drip})$ and inner crust $(\rho>\rho_{\rm drip})$.

The inner crust extends up to the boundary with a core
at a density $\simeq \rho_0$, where nuclei merge into uniform nuclear matter.
Schematic picture of the particle distribution in
neutron star matter in the inner crust is shown in Fig.\ \ref{fig_wscell}(b).
Unlike the supernova matter in the same density region shown in 
Fig.\ \ref{fig_wscell}(a),
dripped neutrons exist outside the neutron-rich nuclei.
Since there are no trapped degenerate neutrinos, 
neutron star matter is much more neutron rich than supernova matter;
typical value of the proton fraction $x$ in the deep region 
of the inner crust and in the core is $x\lesssim 0.1$.

Nuclear pasta phases are expected to exist in the 
bottom layer of the crust (thickness is around $100$ m) 
above the crust-core boundary,
which corresponds to the density region of 
$0.2 \lesssim \rho/\rho_0\lesssim 0.5$.
Note that if the pasta phases exist in neutron stars,
they can occupy half of the mass of the crust \cite{Lorenz93}.

\subsection{Astrophysical implications \label{sect_implication}}

Since the total amount of the pasta phases can be substantial
both in supernova cores and neutron star crusts,
pasta phases can affect various astrophysical phenomena
related to supernovae and neutron stars.
Here, let us briefly see the effects of the pasta phases.

In the collapse phase of a supernova, the mean free path of
neutrinos in the core is reduced mainly by the coherent scattering 
from nuclei via a weak neutral current \cite{Sato75}.  
Suppose a neutrino
is scattered by an isolated nucleus of mass number $A$, the neutrino
is coherently scattered by all the nucleons in the nucleus and the
cross section $\sigma$ is enhanced to be proportional to $A^2$ 
(cf. in the case of incoherent scattering, $\sigma\propto A$) 
provided the neutrino wavelength is much longer than the radius of the nucleus
\cite{Freedman74}.
In the case of the neutrino scattering in matter,
the cross section for coherent scattering is approximated to be
proportional to the static structure factor $S_{nn}(q)$ of neutrons,
where $q$ is the wave number of momentum transfer 
\cite{Horowitz,Horowitz2,opacity}.
Thus existence of the pasta nuclei instead of uniform nuclear matter
or spherical nuclei should strongly modify the neutrino cross section
\cite{Watanabe01}.
This would affect the onset and the duration of neutrino trapping,
and the value of the lepton fraction during this period.

Regarding the pasta phases in neutron stars, we note that the 
crust is located close to the surface of the star (see Fig.\ \ref{fig_ns}).
Therefore, the property of the crust naturally 
influences on many observed phenomena.
Especially, cooling of the neutron star by neutrino emission
is considered to be affected by existence of the pasta phases.

It is well known that 
the most efficient cooling process is the direct URCA process:
$n\rightarrow p+e^-+\bar{\nu}_e$ and
$p+e^-\rightarrow n+\nu_e\,$, but
in neutron star matter, which is degenerate and neutron rich,
this process is strongly
suppressed by the energy and momentum conservation laws.
However, it has been pointed out \cite{Lorenz93} and shown \cite{Gusakov04} 
that the direct URCA process can be open in the pasta phases
with cylindrical bubbles and spherical bubbles
because protons and neutrons move in a periodic
potential created by inhomogeneous nuclear structures
so that they can get lattice momenta needed to satisfy
the momentum conservation (umklapp process).
Cooling simulations of Ref.\ \cite{Gusakov04} show that 
the URCA process in the pasta phases could significantly 
accelerate the cooling of low-mass $(M\simeq 1.35 M_\odot)$ neutron stars.

The elastic property of the pasta phases is very different from that
of the crystalline solid; it is rather similar to that of the liquid
crystal \cite{liquidcrystal}. 
For example, in the pasta phases with rod-like nuclei and
slab-like nuclei --- analogous to the columnar and smectic $A$ phases of
the liquid crystal, respectively --- there are directions in which the
system is translationally invariant and any restoring forces do not act.
Therefore, if the pasta phases exist
between a crystalline solid in the crust and a liquid core,
properties of the crust-core boundary should be strongly modified,
and consequently phenomena related to the crust-core boundary
would be affected.
The $r$-mode instability in rotating neutron stars
is one of the important mechanisms for the gravitational wave radiation
of neutron stars \cite{Andersson98}.
This mechanism accompanies the relative motion between 
the crust and the core and crucially depends on the boundary condition 
between these two regions.
If we neglect the pasta phases and assume that a viscous fluid core is 
bounded by a solid crust, there is a strong shear in the boundary layer 
and the $r$-mode is damped by the viscosity in this region \cite{boundary}.
An existence of the liquid crystal-like pasta phases 
in the crust-core boundary
relaxes the shear and reduces
the viscous damping of the $r$-mode instability \cite{elastwall}.

Related to the elastic property of the pasta phases,
effect of the pasta phases on the torsional shear mode of the crust
has been also studied \cite{Sotani11,Gearheart11}.
Due to the small shear modulus of the pasta phases compared to that of the
crystalline solid, the shear mode frequency can be significantly reduced.

Finally, the pasta phases would have an influence also on 
glitch phenomena of pulsars.
A glitch is a sudden decrease of the pulse period, i.e.,
a sudden spinup of neutron star crusts.
In the crust, (dripped) neutrons are considered to form a superfluid and
their angular momentum is carried by quantized vortices.
On the other hand, nuclei in the crust, 
which mainly consist of a normal fluid, act as pinning centers
for the vortices in the neutron superfluid.
By electromagnetic radiation, the normal component of the crust 
is decelerated and the relative velocity between 
the nuclei and the neutron superfluid increases.
Consequently, a strong Magnus force acts to unpin the pinned vortices.
In the vortex pinning model \cite{Anderson75}, a neutron superfluid whose rotation velocity
is higher than that of nuclei works as a reservoir of the angular momentum.
Glitches are explained as
the angular momentum transfer from the former to the latter
by a catastrophic unpinning of vortices and their migration outward,
which leads to a reduction of the angular velocity of the superfluid.
It is easy to imagine that the structure of nuclei have 
a large influence on the pinning energy of vortices in this scenario.

\section{Nuclear ``pasta'' \label{sect_pasta}}

\subsection{First-order phase transitions and inhomogeneous structures \label{sect_first}}

Nuclear matter with pasta structures can be regarded as a
structured mixed phase during the first-order liquid-gas phase transition.
Here we briefly explain the basic notion of the mixed phase.

The simplest case of the first order liquid-gas phase transition is
that in a single-component system such as water.  Let us consider its
liquid-gas phase transition by changing density at a constant
temperature.  In this case, the equation of state (EOS) during the
liquid-gas phase transition is obtained by the Maxwell
construction, which gives the phase equilibrium between the liquid
and the gas phases at a constant pressure.

On the other hand, systems with multiple components 
like mixture of water and ethanol undergo a more complex 
phase transition. There, pressure in the mixed phase is no longer constant 
during the phase transition and the Maxwell construction cannot be applied.
To obtain the EOS, equilibrium of partial pressures and
chemical potentials, i.e., the Gibbs conditions, should be solved.

\begin{figure}[tbp]
\begin{center}
\rotatebox{0}{
\resizebox{7cm}{!}
{\includegraphics{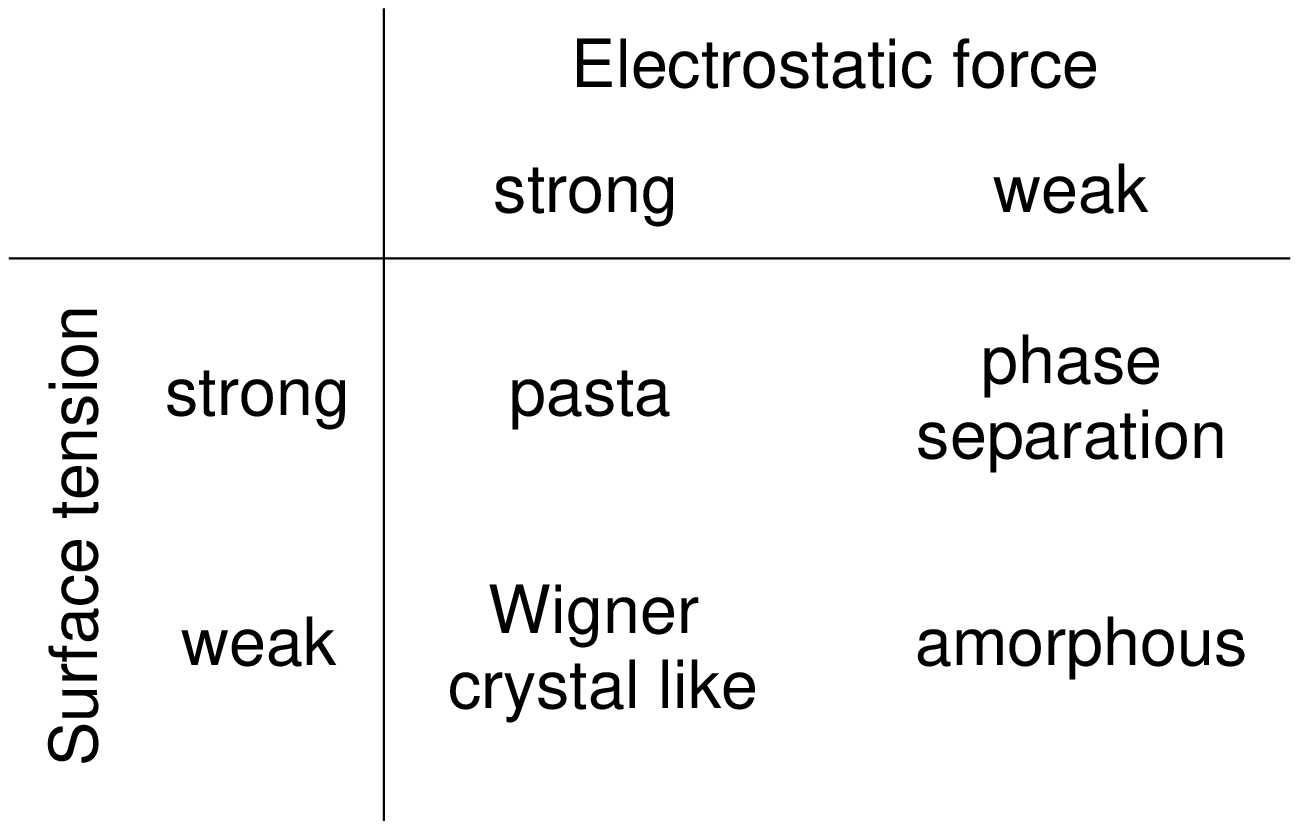}}}
\caption{\label{table_finite}\quad
  Effects of the Coulomb force and the surface tension on the structure
in the mixed phase.
  }
\end{center}
\end{figure}

Nuclear matter, i.e., a mixture of protons and neutrons, 
neutralized by electrons is a system 
with two independent components with electric charge.
Due to the electric charge,
two phases in the mixed phase interact with each other.
Another important factor in the mixed phase of nuclear matter is 
the surface tension of the interface between the liquid phase 
and the gas phase.
In the case of nuclear matter, both the effects of the Coulomb force
and the surface tension are substantial and therefore, 
due to the competition of these two,
rich geometric structures can emerge.
On the other hand, in the previous two cases 
(i.e., water and mixture of water and ethanol), 
while the effect of the surface tension is substantial,
interaction between coexisting phases is negligible.
Thus we have a simple phase separation in these systems 
(see Fig.\ \ref{table_finite}).
If both the effects are weak enough, the structure of mixed phases can be
arbitrary and becomes amorphous
\footnote{If the surface tension is negligible but the effect of the 
Coulomb force is large, the mixed phase takes a Wigner crystal-like structure
in which the phase with charged particles breaks up into a microscopic scale
and form a crystalline lattice.}.

\subsection{Historical overview of the study of nuclear ``pasta'' \label{sect_history}}

The first idea of the phase with non-spherical nuclei was brought out
in a seminal paper by Baym, Bethe, and Pethick in 1971 \cite{Baym}.
They predicted ``nuclei to turn inside out'' 
close to the normal nuclear density: 
Bubbles of neutron gas 
form a crystalline lattice structure in the liquid of nuclear matter.

In early 1980's, Ravenhall {\it et al.} and Hashimoto {\it et al.} 
have independently pointed out the five typical types of the pasta phases
shown in Fig.\ \ref{fig_pasta} can be the ground state of matter.
Using a liquid-drop model, they calculated 
the total energy of the system for the five phases 
and found that the structure of nuclear matter
in the energy minimum state
changes in the sequence from (a) to (e) of Fig.\ \ref{fig_pasta}
with increasing density.

Since these pioneering works, existence of the pasta phases
has been examined by various methods such as 
the liquid-drop model \cite{Oyamatsu84,Lorenz93,Watanabe00,Watanabe01} and
the Thomas-Fermi approximation 
\cite{Wil85,Lassaut87,Oyamatsu93,Sumiyoshi95,Cheng97,Maruyama05,Avancini09}
based on various nuclear forces 
including the Skyrme interactions \cite{Wil85,Lassaut87,Lorenz93}
and the relativistic mean-field models 
\cite{Cheng97,Maruyama05,Avancini09}, etc.
These studies have confirmed that, when the matter is less neutron rich 
(i.e., the average proton fraction $x$ of matter is $0.3\lesssim x \leq 0.5$)
as in the case of supernova matter,
the pasta phases can be the ground state for almost all nuclear models.
As for neutron-rich case such as neutron star matter,
existence of the pasta phases in the ground state has also been
confirmed by many nuclear models, except for those 
(e.g., the relativistic mean-field models \cite{Cheng97,Maruyama05,Avancini09}) 
which give relatively large energy density of neutron matter 
at higher densities of $\gtrsim \rho_0/2$.
Actually, in the case of neutron star matter, 
symmetry energy and its density dependence is
crucial for existence of the pasta phases.
Within the Thomas-Fermi approximation, this issue has been 
studied systematically in Ref.\ \cite{Oyamatsu-Ldependence}.

After the early 2000's,
in addition to the macroscopic liquid-drop model and the semiclassical 
Thomas-Fermi approximation, 
the pasta phases have been studied
by more elaborated calculations based on the Hartree-Fock (+BCS) approximation
with the Skyrme interaction \cite{shell-n,Gogelein07,Newton09}.
There, shell effects, which are absent in the previous calculations
by the liquid-drop model and the Thomas-Fermi approximation,
are naturally incorporated.
Also in these Hartree-Fock calculations, 
existence of the pasta phases has been confirmed 
in both the neutron-rich \cite{Gogelein07} and 
less neutron-rich \cite{Newton09} cases.

Possibility of other exotic nuclear shapes besides the typical
pasta structures has been also discussed \cite{Lassaut87,Nakazato09}.
The double-diamond and the gyroid structures, which actually exist
in the mixture of polymers, are good candidates.
Although these structures have not been found in the ground state of dense matterso far,
the energy difference between the ground state and the phases with these
structures can be very small.
This suggests that these structures might appear at nonzero temperatures.

All of the above studies are based on the static framework and focus
only on the equilibrium state, mainly the ground state.  Thus,
dynamical problems such as the formation process of the pasta phases
have been completely open.  
Besides, the shape of nuclei is assumed in all 
studies of the pasta phases before the early 2000's except for
the work by Williams and Koonin \cite{Wil85}.
Even in their calculation, the size of the simulation box is
so small that it can contain only one period of the pasta structures.
This implicitly imposes a constraint on the nuclear structure, which
prevents the appearance of complex shapes.  In this situation, we
studied the pasta phases using a dynamical framework
\cite{QMD-wata-rapid,QMD-wata,QMD-finiteT-wata,QMD-wata-transition,QMD-sono,QMD-wata-formation}
called the quantum molecular dynamics (QMD) \cite{Aichelin,Peilert,QMD-maru}.
We will discuss QMD in the next section.
Using this framework, we can simulate dynamical
processes in a large size of the simulation box without any
assumptions on the nuclear shape.
This is the main scope of the present article and the results
of our studies will be shown in Section 4.

\section{Quantum Molecular Dynamics \label{sect_qmd}}

\subsection{Microscopic simulation for heavy-ion collisions}

Quantum molecular dynamics (QMD) \cite{Aichelin} 
is one of the microscopic simulation methods
developed for studying intermediate-energy 
($\sim$ 100 MeV per nucleon in the laboratory frame corresponding
to the Fermi energy in the center of mass frame) heavy-ion 
collision processes in the late 1980's.
It describes the time evolution of all the constituent nucleons 
and physical quantities are analyzed using the information of 
each nucleon in the 6$N$-dimensional phase space.

There are several other frameworks of the microscopic simulations used
in the low and intermediate energy regions:
the time-dependent Hartree-Fock (TDHF) and 
kinetic approaches based on the Boltzmann equation such as
the test-particle method of 
the Vlasov equation and that of the Vlasov-Uehling-Uhlenbeck (VUU) equation
\cite{VUU}.
The TDHF deals with the time evolution of the wave function of many-body
system which is approximated by a Slater determinant.
This contains important features of a fermion many-body system such as
the antisymmetrization of the total wave function and the deformation of single-particle
wave functions.
The Vlasov equation is a semi-classical approximation of the TDHF equation
describing the time evolution of the one-body distribution function.
The VUU equation is similar to the Vlasov equation, 
but the two-body (nucleon-nucleon) collision process 
taking account of the Pauli blocking is included.
These frameworks are basically mean-field theories and 
cannot describe processes with higher-order correlations such as
the formation of fragments.

QMD model has been proposed by Aichelin and St\"ocker \cite{Aichelin}
\footnote{
Note that the QMD discussed here is different from 
``quantum molecular dynamics'' used in material science.
}. 
The main quantum aspect included in QMD is 
the stochastic nucleon-nucleon collision process
taking account of the Pauli-blocking
according to the phase-space occupation number.
Technically, QMD simulation is similar to the test-particle method
of the VUU equation (VUU simulation). 
However, unlike VUU simulations, 
QMD can well describe fragment formation processes
including the particle spectrum and the fragment-mass distribution.
It is known that the VUU simulation reduces to the QMD simulation when 
the number of test particles per nucleon is reduced to one.
With decreasing the number of test particles, both
fluctuation in the density distribution of a single event 
and event-by-event fluctuation become larger.
On the other view point, VUU simulation can be regarded as
an ensemble average of many QMD simulation events.
This averaging washes out the fluctuation among collision events,
and consequently, the fluctuation in the density distribution. 
In the 1990's, 
intermediate energy heavy-ion collisions have been intensively analyzed
using QMD simulations.

\subsection{Antisymmetrization of the wave function in molecular dynamics}

Although early version of QMD was successful in the intermediate energy region,
its applicability was not guaranteed in the low energy region.
Since the total wave function in QMD is not antisymmetrized,
early QMD cannot describe the ground state of nuclei properly.
In the energy minimum state of this model, 
all nucleons in nuclei have zero momentum and are deeply bound.
To remedy this problem practically, nuclei with higher internal energy
in which nucleons have nonzero velocities are often used 
in early QMD simulations.
However, these nuclei can spontaneously evaporate nucleons by
decaying into the lower energy state
and such spurious evaporation becomes a serious problem
for simulations of low-energy phenomena with longer time scales.

The fermionic molecular dynamics (FMD) \cite{FMD} 
and the antisymmetrized molecular dynamics (AMD) \cite{AMD}
were developed to resolve the above problem of QMD. 
They use a Slater determinant for the total wave function and
can describe the ground state of nuclei taking account of the Fermi statistics 
in a proper manner.
FMD and AMD successfully describe 
nuclear structures as well as 
dynamical processes in low-energy heavy-ion collisions.
The problem is, however, a huge amount of computing cost
to solve the equations of motion of FMD and AMD,
which is proportional to the fourth power of the particle number $N$
(cf. $\propto N^2$ for QMD).
Thus the use of FMD and AMD has been limited to small systems 
with the total number of particles up to a few hundreds.

\subsection{Pauli potential in QMD}

%
%
In this situation, a phenomenological way to mimic the Pauli principle
using a repulsive two-body potential \cite{Wilets} was introduced in QMD
\cite{Dorso,Peilert}.  This repulsive potential, so-called the Pauli potential,
is a function of the distance not only in the coordinate space 
but also in the momentum space, and acts between nucleons with the same spin
and isospin so that it prevents those particles from 
coming close in the phase space.
Due to the momentum dependence of the Pauli potential,
constituent nucleons have non-zero values of the momentum in the ground state of a nucleus
keeping their velocities at zero; thus the above mentioned
spurious evaporation is avoided.

A typical form of the Pauli potential $V_{\rm Pauli}$ is the Gaussian form,
\begin{equation}
  V_{\rm Pauli}({\bf R}_i,{\bf R}_j; {\bf P}_i,{\bf P}_j)
  =C_{\rm P} \left( \frac{\hbar}{q_{0} p_{0}} \right)^{3}\
  \exp{\left[ -\frac{({\bf R}_i-{\bf R}_j)^2}{2q_0^2} 
    -\frac{({\bf P}_i-{\bf P}_j)^2}{2p_0^2} \right]}\ ,\label{eq_pauli}
\end{equation}
where ${\bf R}_i$ and ${\bf P}_i$ are the position
and momentum of $i$th nucleon, 
$C_{\rm P}$ is the strength, and $q_0$ and $p_0$ are 
the width in the coordinate and the momentum space, respectively.
The widths $q_{0}$ and $p_{0}$ determine
the range of the phase space distance in which 
the exchange repulsion acts.
Therefore, their product should be comparable to 
the volume element of the phase space: $q_{0} \cdot p_{0} \sim h$.
All these parameters 
in the Pauli potential are determined by
fitting the kinetic energy of the non-interacting Fermi gas
\footnote{
The extra potential energy due to the Pauli potential itself
is renormalized into the nuclear potential energies.}.

Unlike the early version of QMD,
QMD with the Pauli potential can describe nuclear matter and finite nuclei
in the ground state by appropriately setting the nuclear potentials.
For example, in our model of Ref.\ \cite{QMD-maru},
nuclear potentials are determined to reproduce the energy, 
($-16$ MeV per nucleon) and the density ($\rho_0=0.165$ fm$^{-3}$) 
of nuclear matter in the ground state, 
binding energies of nuclei, 
and the observed incident-energy dependence of the proton-nucleus potential.
The symmetry energy at the normal nuclear density $\rho_0$ is $34.6$ MeV.
Since there is still an ambiguity in the stiffness of nuclear matter,
we provide three sets of parameters giving the incompressibility 
$K=210$, 280, and 380 MeV.

\section{Recent progress of the study of the pasta phases --- formation of the pasta phases \label{sect_ourwork}}

Since most of the previous works about the pasta phases 
assumed the shape of nuclei within a static framework, 
a fundamental problem whether or not the pasta phases are actually
formed in young neutron stars in the cooling process and
supernova cores in the stage of the gravitational collapse was unclear.
In this situation, we have approached the above questions
using the dynamical framework of QMD and have obtained the following results:
1) Pasta phases can be formed from hot uniform nuclear matter by decreasing temperature.
2) Pasta phase with rod-like nuclei can be formed from a bcc lattice
by compression.
In the present section, we explain our method and results.

\subsection{Formation of nuclear pasta from hot nuclear matter\label{sect_qmdns}}

Now we explain our works where we studied the 
formation of the pasta phases by cooling hot nuclear matter
\cite{QMD-wata-rapid,QMD-wata,QMD-finiteT-wata,QMD-sono}.
This process corresponds to the formation of the pasta phases
in young neutron stars in the cooling process.

In our simulations, we considered a system with 
neutrons, protons, and electrons in a cubic box with periodic boundary
conditions.  The system is not magnetically polarized,
i.e., it contains equal numbers of protons (and neutrons) with 
spin up and spin down.
Relativistic degenerate electrons which ensure charge neutrality can
be regarded as a uniform background because electron screening is
negligibly small at relevant densities around $\rho_0$
\cite{screening,Maruyama05}.
Consequently, one must take account of the long-range nature of
the Coulomb interaction. 
We calculate the Coulomb interaction by the Ewald summation method,
which enables us to sum up the contributions of long-range interactions
in a system with periodic boundary conditions efficiently.
For nuclear interaction, we used the QMD Hamiltonian of 
Ref.\ \cite{QMD-maru} with the
standard medium-EOS parameter set
and that of Ref.\ \cite{QMD-chika}.
The qualitative results are the same for the both models.

First we prepare a uniform hot nucleon gas at a temperature $T\sim
20$ MeV for various densities.  Then using the frictional relaxation
method, we cool it down slowly for $O(10^3-10^4)$ fm$/c$ 
to keep the quasi-thermal equilibrium
throughout the cooling process (the density is kept constant 
during the process).

\begin{figure}[tbp]
\begin{center}
\resizebox{14cm}{!}
{\includegraphics{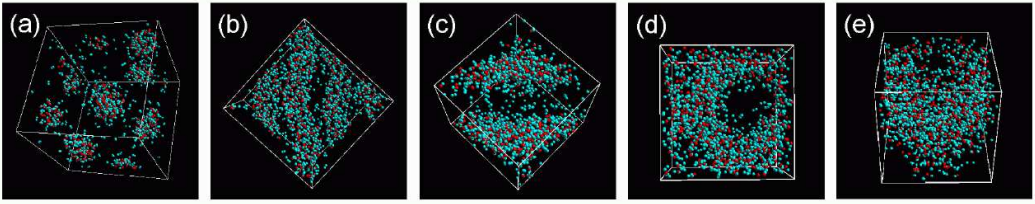}}
\caption{\label{fig_x0.3_t0}
  Nucleon distributions of the pasta phases for $x=0.3$ at $T\simeq 0$.
  The total number of nucleons in the simulation is $2048$ ($614$ protons
  and $1434$ neutrons).
  The red particles show protons and the green ones neutrons.
  Each panel shows the pasta phase with (a) spherical nuclei ($0.1\rho_0$),
  (b) rod-like nuclei ($0.18\rho_0$), (c) slab-like nuclei ($0.35\rho_0$),
  (d) rod-like bubbles ($0.5\rho_0$), and (e) spherical bubbles ($0.55\rho_0$).
  This figure is adapted from Ref.\ \cite{QMD-wata}.
  }
\end{center}
\end{figure}

In Fig.\ \ref{fig_x0.3_t0}, we show the resulting nucleon distributions
at $T\simeq 0$ for various densities of $\lsim \rho_0$.
Here we set the proton fraction of matter $x=0.3$;
the total number of particles in this simulation is $2048$ ($614$ protons
and $1434$ neutrons).
We see that all the typical pasta phases, such as those with spherical nuclei
(a), rod-like nuclei (b), slab-like nuclei (c), rod-like bubbles (d), 
and spherical bubbles (e) have been obtained.
Note that, in this simulation, 
we did not impose any assumption of the nuclear shape,
and these exotic structures were formed spontaneously.

\begin{figure}[tbp]
\begin{center}
\resizebox{14cm}{!}
{\includegraphics{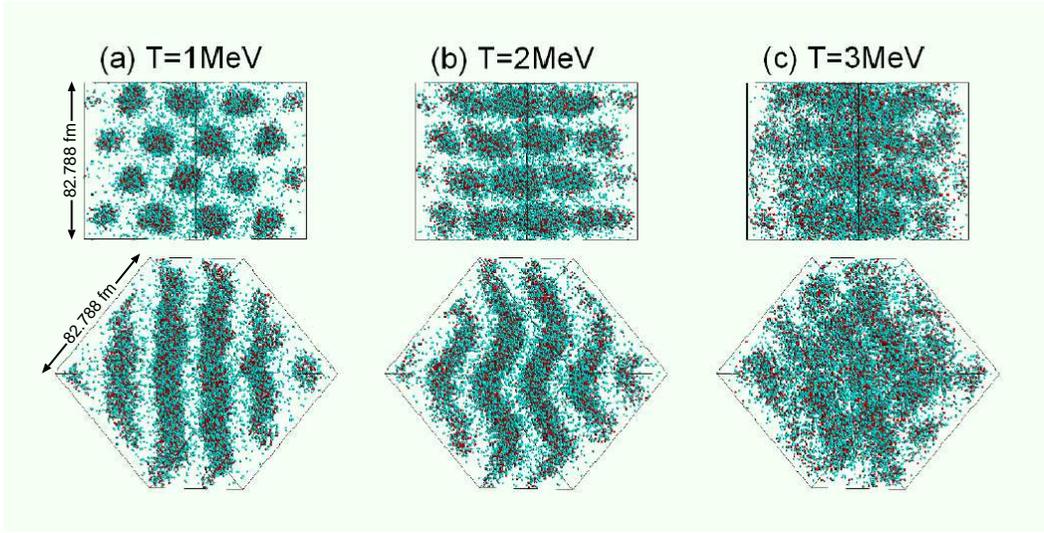}}
\caption{\label{fig_rod}
  Nucleon distributions at $T=1$, 2, and 3 MeV 
  for $x=0.3$, $\rho=0.175\rho_{0}$, where the phase with rod-like nuclei
  is obtained at zero temperature.
  The total number of nucleons in this simulation is $16384$ 
  ($4915$ protons and $11469$ neutrons).
  The upper panels show top views along the axis of the rod-like nuclei 
  at $T = 0$, and the lower ones show side views.
  The red particles show protons and green ones neutrons.
  This figure is taken from Ref.\ \cite{QMD-finiteT-wata}.
  }
\end{center}
\end{figure}

\begin{figure}[tbp]
\begin{center}
\resizebox{14cm}{!}
{\includegraphics{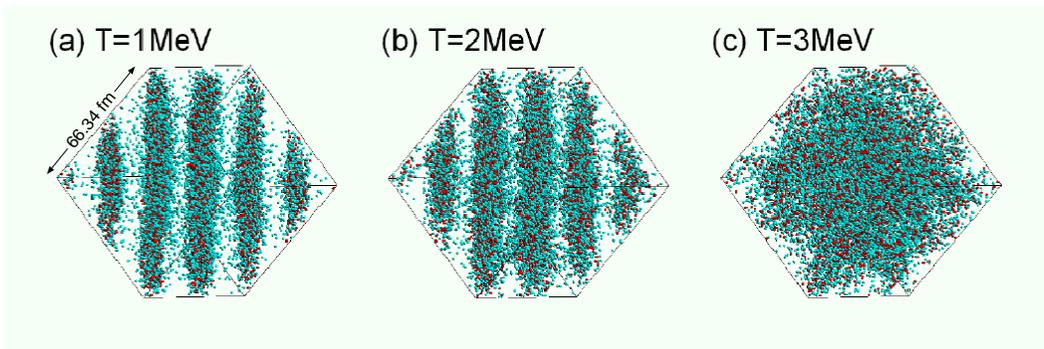}}
\caption{\label{fig_slab}
  The same as Fig.\ \ref{fig_rod} for $x=0.3$, $\rho=0.34\rho_{0}$,
  where the phase with slab-like nuclei is obtained at zero temperature.
  These figures are shown in the direction parallel
  to the plane of the slab-like nuclei at $T=0$.
  This figure is taken from Ref.\ \cite{QMD-finiteT-wata}.
  }
\end{center}
\end{figure}

In Figs.\ \ref{fig_rod} and \ref{fig_slab}, we show the
nucleon distributions at non-zero temperatures
for $x=0.3$ and $\rho=0.175 \rho_0$ and $0.34 \rho_0$, respectively.
At $T=0$, we have obtained the phase with rod-like (slab-like) nuclei 
at the former (latter) density.
In these figures, we can see how the pasta phases are formed 
from a hot uniform nucleon gas by decreasing temperature. 
At relevant densities, the liquid-gas phase separation occurs
at $T\sim 5$ MeV and the density inhomogeneity becomes significant
at $T\sim 3$ MeV.  At $T\simeq 2$ MeV, although 
the surface diffuseness of nuclei and the 
fluctuation of nuclear shape are large and there are still
many evaporated nucleons, clustering of nucleons develops 
and the nuclear shape becomes recognizable.
By further decreasing temperature, surface diffuseness,
global fluctuation of nuclear shape, and the number of evaporated nucleons
get smaller and finally clear rod-like and slab-like nuclei can be observed
at $T\lsim 1$ MeV.

The proton fraction of the above results is $x=0.3$, which is 
higher than the typical value in the neutron star crust, $\lsim 0.1$.
In Ref.\ \cite{QMD-wata}, we also studied the case of
$x = 0.1$ as a more realistic condition for neutron star matter.
There, as well as the phase with spherical nuclei (at $\rho\lsim 0.2\rho_0$),
the pasta phase with rod-like nuclei were obtained at 
$\simeq 0.2\rho_0$ by cooling down hot uniform nucleon gas
(in this case, we took $\sim 10^4$ fm$/c$ to cool from $10$ to 0 MeV).
These results strongly support that the pasta phases
can be formed dynamically in the cooling process of 
young neutron stars.


\subsection{Formation of nuclear pasta by compression\label{sect_qmdsn}}

\subsubsection{Generally accepted scenario based on fission instability}

Formation process of the pasta phases by compression of matter
in collapsing supernova cores is a very non-trivial problem.
Since formation of the pasta phases from a bcc lattice of spherical nuclei 
must be accompanied by dynamical and drastic changes of the nuclear structure,
the fundamental question whether or not the pasta phases 
are formed in supernova cores was a long standing question.

A generally accepted scenario is that the pasta formation is
triggered by fission instability of nuclei \cite{Pethick95}: 
Namely, above some
critical density, at which the volume fraction 
$u$ occupied by nuclei exceeds $1/8$,
the effect of the Coulomb repulsion between protons,
which tends to deform the nucleus, dominates over that of the surface
tension of nucleus to make the nucleus spherical.
Consequently, nuclei are expected to undergo a quadrupole deformation
and they would ``eventually join up to form string-like structures'' 
\cite{Pethick95}.  
Below, we explain the fission instability in detail
and derive the critical value of $u=1/8$.

We consider a nucleus with the mass number $A$, the proton number $Z$, and
the radius $r_{\rm N}$.  We approximate that the density inside the nucleus
is constant and the self-Coulomb energy of the nucleus reads
$E_{\rm Coul}^{(0)}\equiv (3/5)(Z^2e^2/r_{\rm N})$.
A condition for vanishing of the fission barrier with respect to 
a quadrupole deformation is given by
\begin{equation}
  E_{\rm Coul}^{(0)} \geq 2 E_{\rm surf}\quad \mbox{(Bohr-Wheeler's condition)},
\label{eq_bohr_wheeler}
\end{equation}
where $E_{\rm surf}$ is the surface energy of the nucleus.
Physical meaning of this equation is that, 
when the Coulomb energy is sufficiently larger than the surface energy,
the energy gain of the Coulomb repulsion due to the fission exceeds 
the energy cost of the surface tension by increase of the surface area.
Now we consider a lattice of nuclei, where the charge of protons
are neutralized by the background electrons.
Within the Wigner-Seitz approximation, 
where the actual unit cell is replaced 
by a spherical cell of the same volume,
the Coulomb energy $E_{\rm Coul}$ per
nucleus of the lattice is given by
\begin{equation}
  E_{\rm Coul}\simeq E_{\rm Coul}^{(0)}\left(1-\frac{3}{2}u^{1/3}\right).
\label{eq_ec}
\end{equation}
Since $r_{\rm N}\sim A^{1/3}$, the Coulomb and the surface energies per
nucleon scale as $E_{\rm Coul}/A\sim A^{2/3}$ and 
$E_{\rm surf}/A \sim A^{-1/3}$ for fixed $Z/A$ and $u$.
Therefore, the energy minimization with respect to $A$, 
$\partial_A(E_{\rm Coul}/A + E_{\rm surf}/A)=0$, leads to
$E_{\rm surf}=2E_{\rm Coul}$.  Using this equation and Eq.~(\ref{eq_ec}),
Bohr-Wheeler's condition (\ref{eq_bohr_wheeler}) reads
$u\geq 1/8$.

Although the above discussion provides a clear physical insight,
we should keep in mind that Bohr-Wheeler's condition 
(\ref{eq_bohr_wheeler}) is derived for an isolated nucleus (in vacuum).
In the actual situation of supernova matter and neutron star matter,
there are background electrons which reduce
the local net charge density inside nuclei. Thus the condition
for the fission instability should be modified from Bohr-Wheeler's one.
This point has been studied by Brandt \cite{Brandt} (and revisited by 
Ref.\ \cite{Burvenich07}) and 
he has shown that the fission instability is suppressed
by the existence of background electrons which decrease
the Coulomb energy of nuclei.
This result poses a doubt about the generally accepted scenario
of the formation of the pasta phases based on the fission instability.

\subsubsection{Simulations and results}

\begin{figure}[tbp]
\begin{center}
\resizebox{11cm}{!}
{\includegraphics{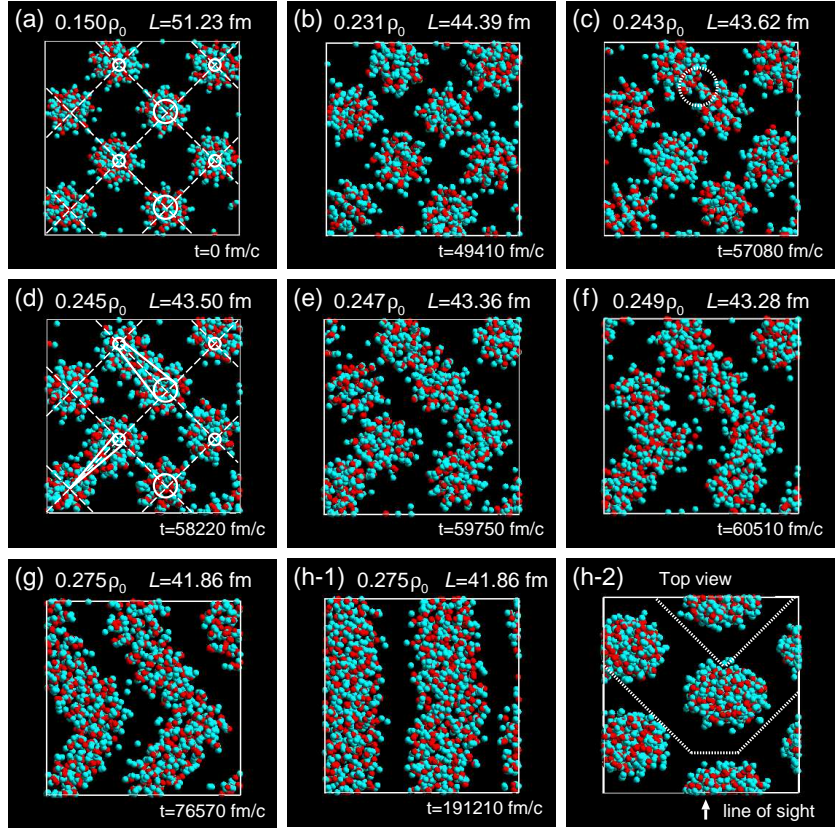}}
\caption{\label{fig_snapshot}
Snapshots of the formation process
of the pasta phase with rod-like nuclei from the bcc lattice
of spherical nuclei by compression of matter.
The red particles show protons and the green ones neutrons.
In panels (a)-(g)
and (h-1), nucleons in a limited region [surrounded by the dotted
lines in panel (h-2)] are shown for visibility.  The vertices of the
dashed lines in panels (a) and (d) show the equilibrium positions of
nuclei in the bcc lattice and their positions in the direction of the
line of sight are indicated by the size of the circles: vertices with
a large circle, with a small circle, and those without a circle are in
the first, second, and third lattice plane, respectively.
The dotted circle in panel (c) shows the first pair of
nuclei start to touch.
The solid lines in panel (d) represent the direction of the two elongated
nuclei: they take zigzag configuration.  
The box sizes are rescaled to be equal in the figures.
This figure is taken from Ref.\ \cite{QMD-wata-formation}.
}
\end{center}
\end{figure}

To solve the above problem,
we simulated the compression of the bcc lattice 
of spherical nuclei in the collapse of supernova cores using QMD
\cite{QMD-wata-formation}.
Here
we used the QMD Hamiltonian of Ref.\ \cite{QMD-maru}
with the standard medium-EOS parameter set
as in the previous section.
Our simulations were carried out for 
the proton fraction $x\simeq 0.39$ and
the total number of nucleons $N=3328$
(with 1312 protons and 2016 neutrons).

In Fig.\ \ref{fig_snapshot}, we show the snapshots of the formation
process of the pasta phase in adiabatic compression.  Starting
from an initial condition at $\rho=0.15 \rho_0$ and 
$T=0.25$ MeV [Fig.\ \ref{fig_snapshot}(a)],
we increased the density by changing the box size $L$ slowly (the
particle positions were rescaled at the same time).  Here the average
rate of the compression was $\gtrsim \mathcal{O}(10^{-6})\ \rho_0/($fm$/c)$ 
yielding the time scale of $\gtrsim 10^{5}$ fm$/c$ to reach the typical density
region of the phase with rod-like nuclei.
While this time scale is, of course, much smaller than the actual time scale
of the collapse, it is 
much larger than that of the change of
nuclear shape (e.g., $\sim 1000$ fm$/c$ for the nuclear fission) and thus the
dynamics observed in our simulation should be determined by the
intrinsic physical properties of the system, not by the density change
applied externally.

At $t\simeq 57080$ fm$/c$ and $\rho\simeq 0.243 \rho_0$ in the
compression process [Fig.\ \ref{fig_snapshot}(c)], the first pair of
two nearest-neighbor nuclei started to touch and fused (dotted
circle), and then formed an elongated nucleus.  After multiple pairs
of nuclei became such elongated ones, we observed a zigzag structure
[Fig.\ \ref{fig_snapshot}(d)].  Then these elongated nuclei stuck
together [see Figs.\ \ref{fig_snapshot}(e) and (f)], and all the
nuclei fused to form rod-like nuclei at $t\lsim 72700$ fm$/c$ and
$\rho\lsim 0.267\rho_0$.  At $t=76570$ and $\rho=0.275\rho_0$
[Fig.\ \ref{fig_snapshot}(g)], we stopped the compression; the
temperature was $\simeq 0.5$ MeV at this point.  Finally, we obtained a
triangular lattice of rod-like nuclei after relaxation
[Figs.\ \ref{fig_snapshot}(h-1) and (h-2)].
From the start of the structural transition process triggered by the fusion 
of the first pair of nuclei, 
the transition process completed within the time scale of $O(10^5)$ fm$/c$.

Note that before nuclei deformed to be elongated due to the fission
instability, they stuck together keeping their spherical shape [see
Fig.\ \ref{fig_snapshot}(c)].  Our simulation shows that the pasta
phases are formed without undergoing fission instability.
Besides, in the middle of the transition process, pair of spherical
nuclei got closer to fuse in a way such that the resulting elongated
nuclei took a zigzag configuration 
and then they further connected to
form wavy rod-like nuclei.  This process is very different from the
generally accepted scenario based on fission instability.

\begin{figure}[tbp]
\begin{center}
\resizebox{10cm}{!}
{\includegraphics{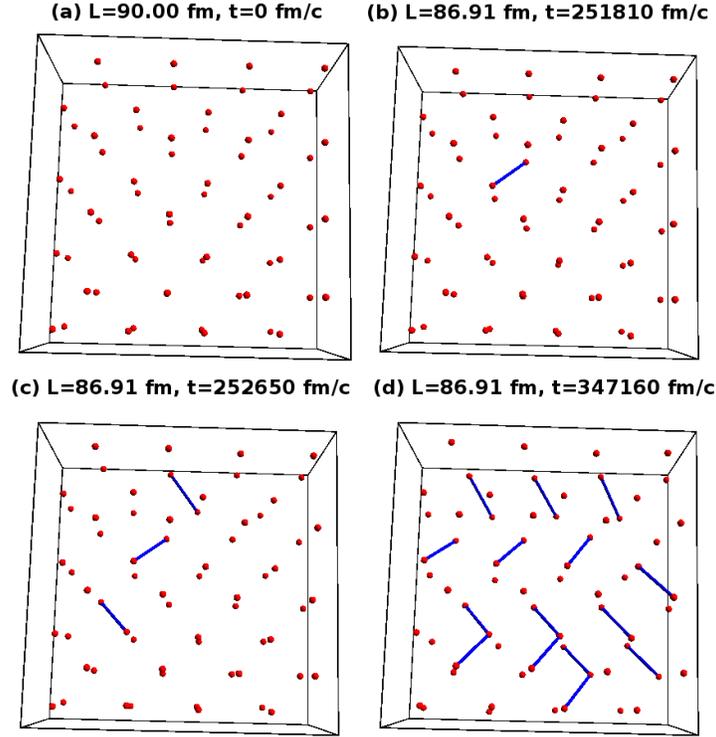}}
\caption{\label{fig_simplemodel} Snapshots of the simulation of
compression at $T < 0.01$ MeV using the simplified model.  The red
particles show the centers of mass of nuclei and the nuclei within the
distance less than $0.89 r_{\mathrm{nn}}^{(0)}$ are connected by a
blue line.  Here, $r_{\mathrm{nn}}^{(0)}$ is the distance between
nearest neighbor site of a bcc lattice for instantaneous density.
Nuclei and connections within only two lattice planes
normal to the line of sight are shown.  This figure is taken from
Ref.\ \cite{QMD-wata-formation}.
}
\end{center}
\end{figure}

Let us now discuss the mechanism of the formation of the zigzag
structure using a simplified model.  Since nuclei start to connect
before they are deformed, it is reasonable to treat a nucleus as a
sphere and incorporate only its center-of-mass degrees of freedom.
When the nearest neighbor nuclei are so close that the tails of their
density profile overlap with each other, net attractive interaction
between these nuclei starts to act due to the interaction between
nucleons in different nuclei in the overlapping surface region.  We
thus consider a minimal model in which each nucleus is treated as a
point charged particle interacting through the Coulomb potential and
a potential of the Woods-Saxon form which describes the finite size
of nuclei and models the internuclear attraction.

With this model, we performed compression of a bcc lattice with 128
nuclei of $^{208}{\rm Pb}$, which corresponds to 8 times larger system
than that of the QMD simulation.  We show the snapshots of this
simulation in Fig.\ \ref{fig_simplemodel}.  In the situation of
Fig.~\ref{fig_simplemodel}(b), the first pair of nuclei started to get
closer and then we stopped the compression and relaxed the system.  We
observed pairings in a zigzag configuration around the first pair
[Fig.~\ref{fig_simplemodel}(c)] and finally we obtained a zigzag structure
[Fig.~\ref{fig_simplemodel}(d)] similar to the one observed 
in the QMD simulation (Fig.~\ref{fig_snapshot}).  
This result shows that the internuclear
attraction caused by overlapping of the surface region of the neighboring nuclei
leads to the spontaneous breaking of the bcc lattice.

What happens if the resulting pasta phase with
rod-like nuclei is further compressed?  
This actually occurs in collapsing supernova cores.
To answer this question, we have performed QMD simulations
of the adiabatic compression of the pasta phases with rod-like nuclei
and slab-like nuclei \cite{QMD-wata-transition}.
These simulations show that 
the pasta phase with rod-like (slab-like) nuclei turns into the phase with
slab-like nuclei (rod-like bubbles) by compression.
(From the start of the transition process, 
it completes within $O(10^4)$ fm$/c$.)
According to these results, we can conclude that,
starting from a bcc lattice of spherical nuclei, 
pasta phases with rod-like nuclei, slab-like nuclei, and rod-like bubbles
are formed sequentially by increasing density during the collapse.

\section{Summary and outlook\label{sect_summary}}

Nuclei with exotic structures such as rod-like and slab-like nuclei 
---  nuclear ``pasta'' --- 
are expected to exist in supernova cores and neutron star crusts.
In this article, we have overviewed the study on nuclear pasta phases;
special focus has been given to the recent progress about the formation
process of the pasta phases studied by a molecular dynamics method
called QMD (Quantum Molecular Dynamics).
A great advantage of this method is that
we can simulate dynamical processes in 
inhomogeneous nuclear matter using a large number of nucleons
without any assumptions on the structure of nuclei.
It is in contrast to many previous studies employing a static framework
within the Wigner-Seitz approximation.

Even though the pasta phases were predicted to exist in the ground
state of nuclear matter at subnuclear densities, it was unclear
whether they are actually formed in the cooling process of hot neutron
star crusts (Section \ref{sect_qmdns}) 
and by the compression of matter in the collapse of
supernova cores (Section \ref{sect_qmdsn}).  In such a situation, we have shown that the pasta
phases can be formed in both of the two cases 
using QMD simulations \cite{QMD-wata-rapid,QMD-wata,QMD-finiteT-wata,QMD-wata-transition,QMD-sono,QMD-wata-formation}.

Since the dynamical formation of the pasta phases has been shown,
the next step to be made is to study the detectability of the pasta phases
in astrophysical phenomena.  Here is our proposal of a to-do list.

\begin{itemize}

\item
There have been many studies about various elementary processes 
in the pasta phases and properties of the pasta phases.
Now it is very important to study their effects
on the neutron star as a whole.
Especially, effects of the pasta phases on the cooling curve
and the oscillation of neutron stars are important problems.
Recently, several authors have started to study these topics 
\cite{Gusakov04,Sotani11,Gearheart11},
but further study in this direction is needed.

\item
To study the $r$-mode instability based on the realistic
crust-core boundary taking account of the pasta phases.
Since the damping of the $r$-mode instability strongly depends
on the condition of the crust-core boundary layer \cite{boundary,elastwall},
the existence of the pasta phases should affect whether or not
the $r$-mode instability is efficient for the gravitational wave radiation.

\item
To make an EOS table for core collapse simulations taking account of the
existence of the pasta phases including their effects on the neutrino opacity.
There is an on-going project in this direction \cite{Furusawa11}.
Core collapse simulations incorporating the pasta phases, which
figure out their influence on the supernova explosion, is highly awaited.

\item
It is natural to consider that supernova cores and neutron star crusts
are polycrystalline.  Ultimately, it is important to understand
the macroscopic properties of matter in supernova cores and neutron star crusts
taking account of the polycrystalline structure.
For example, even though the elastic properties of the bcc Coulomb crystal
\cite{Strohmayer91}
and the pasta phases \cite{liquidcrystal} have been studied, 
actual elastic properties of the macroscopic neutron star crust
should be affected by the grain boundaries in the polycrystal
and differ from those of the single crystal.

\item
It is also important to study the properties of nuclear matter at
subnuclear densities in experiments.  One possibility is as follows
\cite{chiba}: By colliding two nuclei at sub-barrier energies, a low
density region would be created between the nuclei at around their
closest approach.  The density of this region can be controlled by
changing the incident energy.  If fragments are formed in this low
density region, they will be emitted in the transverse direction.  The
properties of these fragments such as the mass number and the proton
fraction can be measured, which would tell us the information of low
density nuclear matter.

\end{itemize}

\section*{Acknowledgements}

The studies reported in this article have been done in collaborations with
Hidetaka Sonoda, Katsuhiko Sato, Kenji Yasuoka, and Toshikazu Ebisuzaki.
In these studies we used MDGRAPE-2 and -3 of the 
RIKEN Super Combined Cluster System.
GW acknowledges the Max Planck Society, 
the Korea Ministry of Education, Science and Technology (MEST), 
Gyeongsangbuk-Do, 
and Pohang City for the support of the Independent Junior Research Group 
at the Asia Pacific Center for Theoretical Physics (APCTP).
TM is grateful to Toshitaka Tatsumi and Satoshi Chiba for fruitful discussions
and to APCTP for warm hospitality during his visit.

\label{lastpage-01}

\end{document}